\documentclass[10pt, final, conference, letterpaper]{IEEEtran}
\usepackage[left=.6in,right=.6in,top=.719in,bottom=.99in]{geometry}

\usepackage{amsmath,amssymb,dsfont,stfloats,color,url,bbm}
\usepackage[pdftex]{graphicx}
\usepackage[caption=false,font=footnotesize]{subfig}
\usepackage{mathabx}

\DeclareGraphicsExtensions{.eps,.pdf,.png,.jpg,.gif,.jpeg,.pstex}

\usepackage[noadjust]{cite}
\usepackage{url}

\newcommand*{\Resize}[2]{\resizebox{#1}{!}{$#2$}}%

\newfont{\bbb}{msbm10 scaled 700}

\newcommand{\EEE}{\mbox{\bbb E}}

\newfont{\bb}{msbm10 scaled 1100}
\newcommand{\CC}{\mbox{\bb C}}
\newcommand{\PP}{\mbox{\bb P}}

\newcommand{\EE}{\mbox{\bb E}}

\newcommand{\HH}{\mbox{\bb H}}
\newcommand{\SSS}{\mbox{\bb S}}

\newcommand{\BB}{\mbox{\bb B}}
\renewcommand{\AA}{\mbox{\bb A}}

\newcommand{\yy}{\mathbbm{y}}

\newcommand{\zz}{\mathbbm{z}}
\newcommand{\sss}{\mathbbm{s}}

\newcommand{\hh}{\mathbbm{h}}

\newcommand{\vvv}{\mathbbm{v}}

\newcommand{\av}{{\bf a}}

\newcommand{\hv}{{\bf h}}

\newcommand{\rv}{{\bf r}}
\newcommand{\sv}{{\bf s}}

\newcommand{\wv}{{\bf w}}
\newcommand{\vv}{{\bf v}}
\newcommand{\xv}{{\bf x}}
\newcommand{\yv}{{\bf y}}
\newcommand{\zv}{{\bf z}}
\newcommand{\zerov}{{\bf 0}}

\newcommand{\Dm}{{\bf D}}

\newcommand{\Fm}{{\bf F}}
\newcommand{\Gm}{{\bf G}}

\newcommand{\Id}{{\bf I}}

\newcommand{\Xm}{{\bf X}}
\newcommand{\Ym}{{\bf Y}}
\newcommand{\Zm}{{\bf Z}}

\newcommand{\Cc}{{\cal C}}

\newcommand{\Ec}{{\cal E}}

\newcommand{\Gc}{{\cal G}}

\newcommand{\Nc}{{\cal N}}

\newcommand{\Sc}{{\cal S}}

\newcommand{\Uc}{{\cal U}}

\newcommand{\nuv}{\hbox{\boldmath$\nu$}}

\newcommand{\zetav}{\hbox{\boldmath$\zeta$}}
\newcommand{\phiv}{\hbox{\boldmath$\phi$}}

\newcommand{\xiv}{\hbox{\boldmath$\xi$}}

\newcommand{\Gammam}{\hbox{\boldmath$\Gamma$}}

\newcommand{\Sigmam}{\hbox{\boldmath$\Sigma$}}

\newcommand{\Xim}{\hbox{\boldmath$\Xi$}}

\newcommand{\diag}{{\hbox{diag}}}

\newcommand{\trace}{{\hbox{tr}}}

\newcommand{\eqdef}{\stackrel{\Delta}{=}}

\newcommand{\herm}{{\sf H}}

\newcommand{\SINR}{{\sf SINR}}
\newcommand{\SNR}{{\sf SNR}}

\usepackage{stmaryrd} 

\begin{document}

\setlength{\abovedisplayskip}{1pt}
\setlength{\belowdisplayskip}{1pt}
\setlength{\abovedisplayshortskip}{1pt}
\setlength{\belowdisplayshortskip}{1pt}

\title{The Impact of Subspace-Based Pilot Decontamination in User-Centric Scalable Cell-Free Wireless Networks}

\author{\IEEEauthorblockN{Fabian G\"ottsch\IEEEauthorrefmark{1},
Noboru Osawa\IEEEauthorrefmark{2}, Takeo Ohseki\IEEEauthorrefmark{2}, Kosuke Yamazaki\IEEEauthorrefmark{2}, Giuseppe Caire\IEEEauthorrefmark{1}}
\IEEEauthorblockA{\IEEEauthorrefmark{1}Technical University of Berlin, Germany\\
\IEEEauthorrefmark{2}KDDI Research Inc., Japan\\
Emails: \{fabian.goettsch, caire\}@tu-berlin.de, \{nb-oosawa, ohseki, ko-yamazaki\}@kddi-research.jp}}

\maketitle


\begin{abstract}
We consider a {\em scalable} user-centric  wireless network with dynamic cluster formation as 
defined by Bj\"ornsson and Sanguinetti. 
Several options for scalable uplink (UL) processing are examined including: i) cluster size and SNR threshold criterion for cluster formation; 
ii) UL pilot dimension; iii) local detection and global (per cluster) combining. 
We use a simple model for the channel vector spatial correlation, which captures the fact that the propagation between UEs and RRHs 
is not isotropic. In particular, we define the ideal performance based on ideal but partial CSI, i.e., the CSI that can be estimated 
based on the users to antenna heads cluster connectivity. In practice, CSI is estimated from UL pilots, and therefore it is affected by noise and pilot contamination. 
We show that a very simple subspace projection scheme is able to basically attain the same performance of perfect but partial CSI. 
This points out that the essential information needed to pilot decontamination reduces effectively to the dominant channel subspaces. 
\end{abstract}


\section{Introduction} 

{\em Multiuser MIMO} is arguably one of the key transformative ideas that have shaped the last 15 years of theoretical research and  
eventually made a very significant impact on actual system design, since the first information theoretical break through of
Caire and Shamai \cite{caire2003achievable}, to the provisions in recent wireless standards \cite{3gpp38211}. 
A successful related concept is Marzetta's {\em massive MIMO} \cite{marzetta2010noncooperative}. 
This is based on the key idea that, thanks to channel reciprocity and TDD operations,  
an arbitrarily large number $M$ of base station (BS) antennas can be trained by a finite number $K$ of user equipments (UE) 
using a finite-dimensional uplink (UL) pilot field $\tau_p \geq K$. 
When using massive MIMO in a large cellular network with per-BS processing, serving $K  \gg \tau_p$ users per cell per channel 
coherence block\footnote{We define a channel coherence block as a ``tile'' of $T$ symbols in the time-frequency domain over which
the fading channel can be considered constant. For the sake of conceptual simplicity, this can be identified as resource block (RB) 
of the underlying PHY protocol.}  
implies that mutually  non-orthogonal pilots are reused across the network yielding  {\em pilot contamination}, which creates a coherent combined term in the
inter-cell interference that does not vanish as $M \rightarrow \infty$ \cite{marzetta2010noncooperative}.  
More recently, a flurry of works advocating the joint processing of spatially distributed infrastructure antennas has appeared. 
This idea can be traced back to the work of Wyner \cite{wyner1994shannon}, and has been ``re-marketed''  several times under different names with slight nuances, 
such as {\em coordinate multipoint} (CoMP),  {\em cloud radio access network} (CRAN), or {\em cell-free massive MIMO}. 
An excellent recent review of this vast literature is given in \cite{9336188}. Advantages of this approach are the mitigation of pathloss and blocking, 
introducing proximity between the remote radio heads (RRHs) and UEs and macro-diversity, and the (obvious) elimination 
of inter-cell interference, by providing a single giant RRH cluster. 
Both points, though, must be carefully discussed. First, deploying a number of RRH much larger than the number of UEs
is practically problematic, very costly, and often infeasible especially for outdoor systems.  
Then, the joint antenna processing across the whole network
does not eliminate the problem of a limited UL pilot dimension $\tau_p \ll K$. 
Finally, global processing, optimization/allocation of pilots and transmit power across the network 
yield a non-scalable architecture. 

Here we adopt the definition of scalability given by Bj\"ornsson and Sanguinetti \cite{9064545}, informally recalled as follows: 
consider a network covering an area $A$ on the plane, with UE and RRH densities $\lambda_u$ and $\lambda_a$, respectively. 
Assume also that the RRHs are connected by a routing-capable fronthaul network to Decentralized Processing Units (DPUs), 
spatially distributed with density  $\lambda_d$.  An architecture is scalable if, as $A \rightarrow \infty$, the complexity of the involved signal processing 
functions and the data rate conveyed at each DPU is $O(1)$ (constant with $A$).  

We consider the realistic case where each RRH has $M$ antennas, and $\lambda_a < \lambda_u < M \lambda_a$. 
Then, the number of ``antenna sites'' is (significantly) less than the number of users $K$ simultaneously 
active on any given RB.  Scalable {\em user-centric} architectures have been recently proposed in several papers
(again see \cite{9336188}), based on dynamic cluster formation, such that every UE is served by a finite-size cluster of RRHs even 
if the network is arbitrarily large. 
In this paper we follow this paradigm and  use a simple model for the 
channel vectors' spatial correlation, capturing the directional propagation between UEs and RRHs. 
We show that a pilot subspace projection scheme is able to approach very closely the performance of perfect CSI, pointing out the essential role
of the channel subspace information.  As a practical remark, we note that in 5GNR two types of UL pilots are specified, 
the demodulation references signals (DMRS) and the sounding reference signals (SRS). 
In this work we assume that the instantaneous CSI is obtained from DMRS pilots, and the subspace information is known. 
In a future work, we will deal with the channel subspace information estimation by exploiting features of the SRS pilots. 
It is worthwhile noticing that the knowledge of the channel subspace information  is less demanding and more robust than 
the full knowledge of the channel covariance matrix, as assumed for example in \cite{9064545,9336188}. 

Throughout this paper, we will use boldface capital letters ($\Xm$) for matrices and boldface small letters ($\xv$) for vectors that contain information of a RRH-UE pair. The composed matrices and vectors that contain information of multiple RRHs and/or UEs are denoted by the blackboard letters $\mbox{\bb X}$ and $\mathbbm{x}$, respectively.

\section{System model}
\begin{figure}[h!]
	\centerline{\includegraphics[trim={140 130 100 82}, clip, width=.45\linewidth]{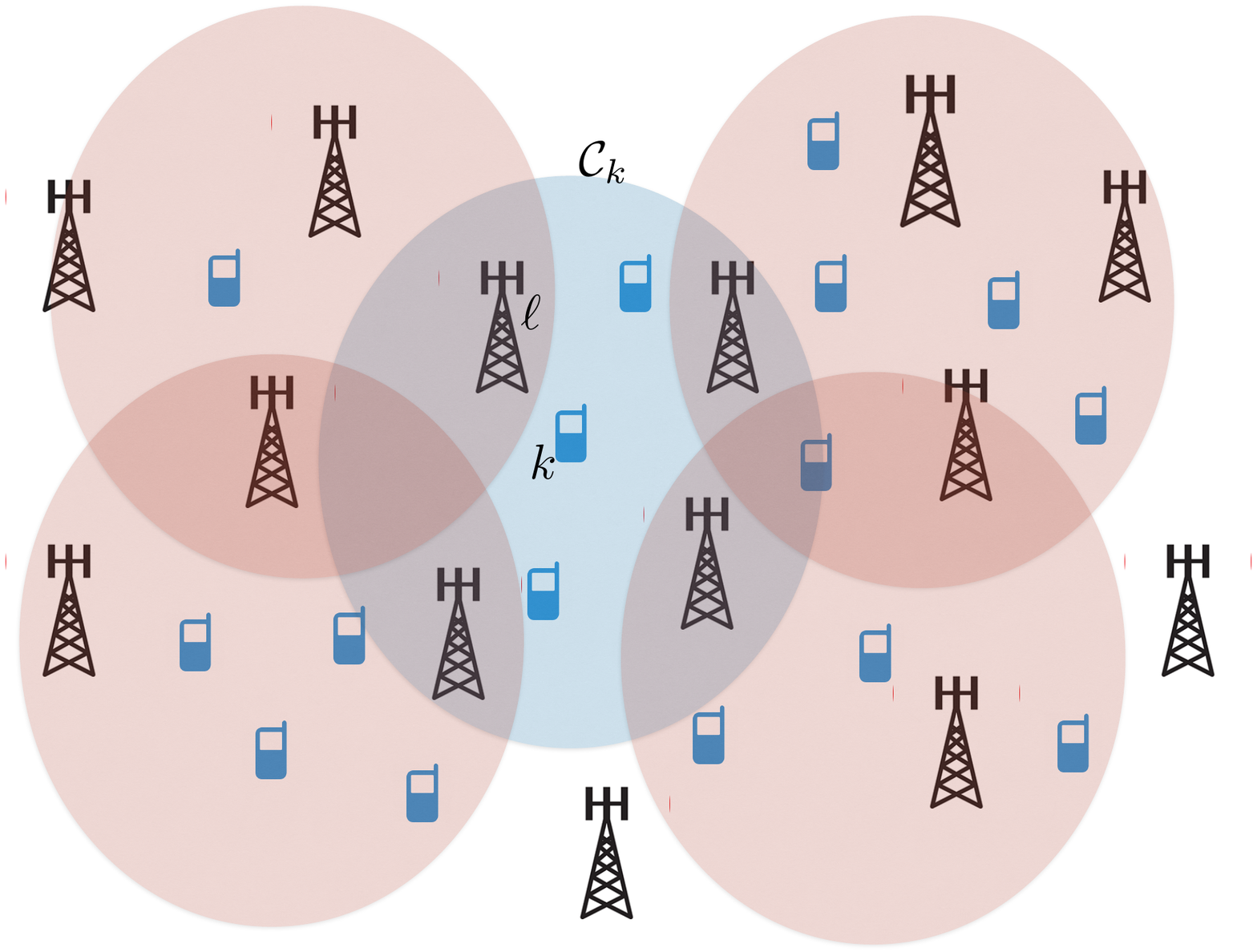} \includegraphics[trim={100 120 140 170}, clip, width=.45\linewidth]{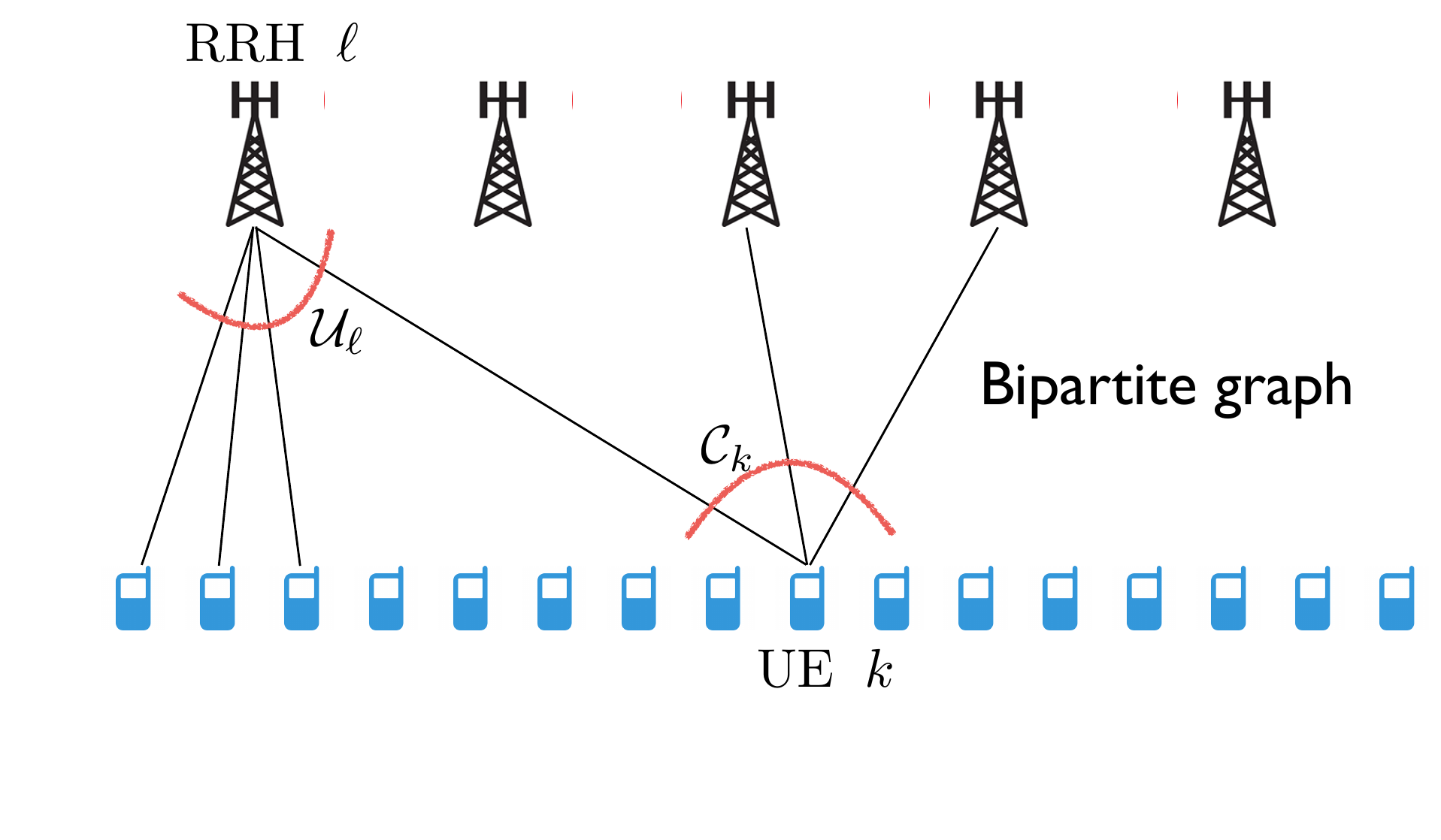}}
	\vspace{-.3cm}
	\caption{\footnotesize An example of dynamic clusters and the UE-RRH association graph. The graph contains a UE-RRH edge $(k,\ell)$ for all 
	$k \in [K]$ and $\ell \in [L]$ such that  $k \in \Uc_\ell$ and $\ell \in \Cc_k$.}
	\label{clusters}
\end{figure}

We consider a cell-free wireless network with $L$ RRHs, each equipped with $M$ antennas, 
and $K$ single-antenna UEs. Both RRHs and UEs are distributed on a squared region on the 2-dimensional plane. 
As a result of the cluster formation process (to be specified later), each UE $k$ is associated with a cluster $\Cc_k \subseteq [L]$ of RRHs 
and each RRH $\ell$ has a set of associated UEs $\Uc_\ell \subseteq [K]$. The UE-RRH association is described by 
a bipartite graph $\Gc$ with two classes of nodes (UEs and RRHs) such that the neighborhood of UE-node $k$ is $\Cc_k$ 
and the neighborhood of RRH-node $\ell$ is $\Uc_\ell$. An example is given in Fig.~\ref{clusters}. 
The set of edges of $\Gc$ is denoted by $\Ec$, i.e., $\Gc = \Gc([L], [K], \Ec)$. 
\vspace{-0.4cm}
\vspace{-0.278cm}

We assume OFDM modulation and assume that the channel in the time-frequency domain follows the
standard block-fading model adopted in countless papers \cite{marzetta2010noncooperative,9336188,9064545},  
where the channel vectors from UEs to RRHs are random but constant over coherence blocks of 
$T = N_{\rm rb} \times N_{\rm sub}$ signal dimensions in the time-frequency domain, where
$N_{\rm rb}$ and $N_{\rm sub}$ indicate  the number of OFDM symbols in time 
and the number of OFDM subcarriers in frequency forming a RB. 

Since all our treatment can be formulated on a per-RB basis, we shall neglect the RB index for the sake of notation simplicity. 
We let $\HH \in \CC^{LM \times K}$ denote the channel matrix between all the $K$ UE antennas and all the $LM$ 
RRHs antennas on a given RB, formed by $M \times 1$ blocks $\hv_{\ell,k}$ in correspondence of the $M$ antennas of RRH $\ell$ 
and UE $k$.  Because of the UL pilot allocation (see later), each RRH $\ell$ only estimates the channel vectors of the users in $\Uc_\ell$.
As a genie-aided best-case, we define the {\em ideal partial CSI} regime
where each RRH $\ell$ has perfect knowledge of the channel vectors $\hv_{\ell,k}$ for $k \in \Uc_\ell$. 
In this regime, the part of the channel matrix $\HH$ known  at the DPU serving cluster $\Cc_k$ 
is denoted by $\HH(\Cc_k)$. This matrix has the same dimensions of $\HH$,  such that the $(\ell, j)$ block 
of dimension $M \times 1$  of $\HH(\Cc_k)$ is equal to $\hv_{\ell,j}$ for all  $(\ell, j) \in \Ec$  and to $\zerov$ otherwise.  

For the individual UE-RRH channels, we consider a simplified directional channel model defined as follows.
Let $\Fm$ denote the $M \times M$ unitary DFT matrix with $(m,n)$-elements
$\Fm_{m,n} = \frac{e^{-j\frac{2\pi}{M} mn}}{\sqrt{M}}$ for  $m, n  = 0,1,\ldots, M-1$. 
Consider the angular support set $\Sc_{\ell,k} \subseteq \{0,\ldots, M-1\}$ 
obtained according to the single ring local scattering model (see \cite{adhikary2013joint}). Then, we let
\begin{equation} 
	\hv_{\ell,k} = \sqrt{\frac{\beta_{\ell,k} M}{|\Sc_{\ell,k}|}}  \Fm_{\ell,k} \nuv_{\ell, k},
\end{equation}
where, using a Matlab-like notation, $\Fm_{\ell,k} \eqdef \Fm(: , \Sc_{\ell,k})$ denotes the tall unitary matrix obtained by selecting the columns 
of $\Fm$ corresponding to the index set $\Sc_{\ell,k}$, $\beta_{\ell,k}$ is a large scale fading coefficient (LSFC) including distance-dependent 
pathloss, blocking effects, and shadowing, and  $\nuv_{\ell,k}$ is an $|\Sc_{\ell,k}| \times 1$ i.i.d. Gaussian vector with components 
$\sim \Cc\Nc(0,1)$. 

\subsection{Cluster formation} \label{sec:cluster_formation}

We assume that $\tau_p$ signal dimension per RB are dedicated to UL pilots (see \cite{3gpp38211}), and define a codebook of $\tau_p$ orthogonal pilots 
sequences.  The UEs transmit with the same power $P^{\rm ue}$,  and we define the system parameter $\SNR \eqdef P^{\rm ue}/N_0$, 
where $N_0$ denotes the noise power spectral density. By the normalization of the channel vectors, the maximum beamforming gain 
averaged over the small scale fading is $\EE[\lVert \frac{M}{|\Sc_{\ell,k}|} \Fm_{\ell,k} \nuv_{\ell, k} \rVert^2] = M$. Therefore, the maximum SNR at the receiver of RRH $\ell$ from UE $k$ is 
$\beta_{\ell,k} M \SNR$.  As in \cite{9064545}, each UE $k$  elects its leading RRH $\ell$ as the RRH 
with the largest channel gain $\beta_{\ell,k}$ (assumed known) among the RRHs with yet a free DMRS pilot and satisfying the 
QoS condition $\beta_{\ell,k} \geq \frac{\eta}{M \SNR}$, where  $\eta > 0$ is a suitable threshold. 
If such RRH is not available, then the UE is declared in outage. In our simulations, the UE to leader RRH association is performed 
in a greedy manner starting from some UE at random. In practice, users join and leave the system according to some
user activity dynamics, and each new UE joining the system is admitted if it can find a leader RRH according to the above conditions. 
After all non-outage UEs $k$ are assigned to their leader RRH $\ell = \ell(k)$ and therefore have a pilot index 
$t = t(k) \in [\tau_p]$, the dynamic cluster $\Cc_k$ for each UE $k$  is formed by enrolling successively all RRH $\ell$ 
listed in order of decreasing LSFC for which i) pilot $t(k)$ is yet free, ii) the condition $\beta_{\ell,k} \geq \frac{\eta}{M \SNR}$ is satisfied,   
and iii) the maximum cluster size $Q$ is not met. As a result, we have that all UEs $k \in \Uc_\ell$ make use of mutually orthogonal UL pilots  
and that  $0 \leq |\Uc_\ell| \leq \tau_p$ and $0 \leq |\Cc_k| \leq Q$.

\subsection{Uplink data transmission} 

The received $LM \times 1$ symbol vector at the $LM$ RRHs' antennas for a single channel use of the UL is given by
\begin{equation} 
	\yy^{\rm ul} = \sqrt{\SNR} \; \HH \sss^{\rm ul}   + \zz^{\rm ul}, \label{ULchannel}
\end{equation}
where $\sss^{\rm ul} \in \CC^{K \times 1}$ is the vector of
of information symbols transmitted by the UEs (zero-mean unit variance and mutually independent random variables) and 
$\zz^{\rm ul}$ is an i.i.d. noise vector with components $\sim \Cc\Nc(0,1)$.  
The goal of cluster $\Cc_k$ is to produce an effective channel observation for symbol $s^{\rm ul}_k$ 
(the $k$-th component of the vector $\sss^{\rm ul}$ from the collectively received signal at the RRHs $\ell \in \Cc_k$).  
We  define the receiver {\em unit norm} vector $\vvv_k \in \CC^{LM \times 1}$ formed by $M \times 1$ blocks
$\vv_{\ell,k} : \ell = 1, \ldots, L$, such that $\vv_{\ell,k} = \zerov$ (the identically zero vector) if $(\ell,k) \notin \Ec$. 
This reflects the fact that only the RRHs in $\Cc_k$ are involved in producing a received observation for the detection of user $k$. 
The non-zero blocks $\vv_{\ell,k} :  \ell \in \Cc_k$ are suitably defined, depending on the receiver combining scheme as examined later. 
The corresponding scalar combined observation for symbol $s^{\rm ul}_k$ is given by 
\begin{eqnarray}
	r^{\rm ul}_k  & = & \vvv_k^\herm \yy^{\rm ul}. 
\end{eqnarray}
%
%
%
For simplicity, we assume that the channel decoder has perfect knowledge of the exact signal to Interference plus noise ratio (SINR) value \vspace{-.05cm}
\begin{eqnarray} 
	\SINR^{\rm ul}_k 
& = & \frac{  |\vvv_k^\herm \hh_k|^2 }{ \SNR^{-1}  + \sum_{j \neq k} |\vvv_k^\herm \hh_j |^2 },  \label{UL-SINR-unitnorm}
\end{eqnarray}
where $\hh_k$ denotes the $k$-th column of $\HH$. The corresponding {\em ergodic} achievable rate is given by 
\begin{equation}
R_k = \EE [ \log ( 1 + \SINR^{\rm ul}_k ) ], \label{ergodic-rate}
\end{equation}
where the expectation is with respect to the small scale fading, while conditioning on the placement of UEs and RRH, and on the cluster formation. 
We refer to (\ref{ergodic-rate}) as the {\em optimistic ergodic rate}.
The use of this performance metric instead of some of the several achievable ergodic rate lower bounds available in the literature 
(e.g., \cite{marzetta2016fundamentals} and discussion therein) 
is justified by the fact that  there is no information theoretic converse proving that such optimistic rates cannot be achieved 
using some form of universal decoder.\footnote{ 
For example, it is well-known that the rate $\frac{1}{2} \log(1 + \gamma)$ per real dimension is achievable
by spherical codes and minimum distance decoding for the channel $y = \sqrt{\gamma} x + z$, with $\EEE[x^2] = \EEE[z^2] = 1$, 
even if $\gamma$ is unknown to the receiver and the noise $z$ is non-Gaussian and uncorrelated with the signal $x$.} 
We do not  claim here the achievability of the optimistic ergodic rates, but we claim that these quantities are ``good enough'' for
comparing the effect of different system parameter configurations on the system performance, without  resorting to 
other bounds that may be indeed overly pessimistic. 
%

\section{UL schemes with ideal partial CSI} \label{sec:ul_perfect_csi}
\subsection{Global Zero-Forcing (GZF)}
\vspace{-0.08cm}
For a given UE $k$ with cluster $\Cc_k$,  we define the set $\Uc(\Cc_k) \eqdef \bigcup_{\ell \in \Cc_k} \Uc_\ell$ of UEs served by at least one RRH in $\Cc_k$. 
Let $\hh_k(\Cc_k)$ denote the $k$-th column of $\HH(\Cc_k)$ and let $\HH_k(\Cc_k)$ denote the residual matrix after deleting the $k$-th column. 
The GZF receiver vector is obtained as follows.  Let $\overline{\hh}_k(\Cc_k) \in \CC^{|\Cc_k|M \times 1}$ and $\overline{\HH}_k(\Cc_k) \in  \CC^{|\Cc_k|M \times (|\Uc(\Cc_k)|-1)}$ the vector and matrix
obtained from $\hh_k(\Cc_k)$ and $\HH_k(\Cc_k)$, respectively, after removing all the $M$-blocks of rows corresponding to 
	RRHs $\ell \notin \Cc_k$ and all the (all-zero) columns corresponding to UEs $k' \notin \Uc(\Cc_k)$. Consider the 
	singular value decomposition (SVD) 
	\begin{equation}
		\overline{\HH}_k(\Cc_k) = \overline{\AA}_k \overline{\SSS}_k \overline{\BB}_k^\herm, 
	\end{equation}
	where the columns of the tall unitary matrix $\overline{\AA}_k$ form an orthonormal basis for the column span of $\overline{\HH}_k(\Cc_k)$, such that 
	the orthogonal projector onto the orthogonal complement of the interference subspace is given by 
$\overline{\PP}_k = \Id - \overline{\AA}_k \overline{\AA}_k^\herm$, 
and define the unit-norm vector 
	\begin{equation} 
	\overline{\vvv}_k = \overline{\PP}_k \overline{\hh}_k(\Cc_k) / \| \overline{\PP}_k \overline{\hh}_k(\Cc_k) \|. 
	\end{equation} 
	Hence, the GZF receiver vector $\vvv_k$ is given by expanding $\overline{\vvv}_k$ by reintroducing the missing blocks of 
	all-zero $M \times 1$ vectors $\zerov$ in correspondence of the RRHs $\ell \notin \Cc_k$. 
	If $M > \tau_p$ (i.e., more antennas than UL data streams), noticing that  $|\Uc_\ell| \leq \tau_p$ (due to the cluster formation rule), we have that
	$|\Uc(\Cc_k)| \leq \tau_p |\Cc_k| < M  |\Cc_k|$.
	Therefore, $\AA_k$ defined before is effectively tall unitary and the global ZF always exists with probability 1 for random/Gaussian 
	user channel vectors. 

	\vspace{-0.1cm}	
	\subsection{Local MRC and MMSE with global combining}
	In this case, each RRH $\ell$ makes use of locally computed receiving vectors $\vv_{\ell,k}$ for 
	its users $k \in \Uc_\ell$. Let $\yv_\ell^{\rm ul}$ denote
	the $M \times 1$ block of $\yy^{\rm ul}$ corresponding to RRH $\ell$. For each $k \in \Uc_\ell$, RRH $\ell$ computes locally
		$r^{\rm ul}_{\ell,k} = \vv_{\ell,k}^\herm \yv_\ell^{\rm ul}$. 
	The symbols $\{r^{\rm ul}_{\ell,k} : k \in \Uc_\ell\}$ are sent to the DPU serving UE $k$, which computes the globally combined symbol 
$r^{\rm ul}_k = \sum_{\ell \in \Cc_k} w^*_{\ell,k} r^{\rm ul}_{\ell,k} = \wv^\herm_k \rv^{\rm ul}_k$,  
where $w_{\ell,k}$ is the combining coefficient of RRH $\ell$ for UE $k$, and $\wv_k$ and $\rv_k$ are vectors formed by stacking $w_{\ell,k}$ and $r^{\rm ul}_{\ell,k}$ of all RRHs $\ell \in \Cc_k$, respectively.
	
	One possible choice for the receiver vector $\vv_{\ell,k}$ is the Maximal Ratio Combining (MRC) receiver, given by 
		$\vv_{\ell,k} = \hv_{\ell,k}$. 
	An alternative choice consists of using a linear MMSE (LMMSE) principle. In this case, we distinguish between the known part of the interference, 
	i.e., the term  $\sum_{j \in \Uc_\ell : j \neq k}  \hv_{\ell,j} s_j^{\rm ul}$, and the unknown part of the interference, 
	i.e., the term  $\sum_{j \notin \Uc_\ell} \hv_{\ell,j} s_j^{\rm ul}$ in $\yv_\ell^{\rm ul}$. 
	The receiver treats the unknown part of the interference plus noise as a white vector with known variance per component. 
	The covariance matrix of this term is given by 
	\begin{align} 
		&\Resize{1\linewidth} {\Xim_\ell = \EE \left [ \left ( \sqrt{\SNR} \sum_{j \notin \Uc_\ell} \hv_{\ell,j} s_j^{\rm ul}  + \zv_\ell^{\rm ul} \right )
		\left ( \sqrt{\SNR} \sum_{j \notin \Uc_\ell} \hv_{\ell,j} s_j^{\rm ul}  + \zv_\ell^{\rm ul} \right )^\herm \right ]} \nonumber \\ 
		& \hspace{.32cm} = \Id +  \sum_{j \notin \Uc_\ell} \frac{\beta_{\ell,j} M \SNR }{|\Sc_{\ell,j}|} \Fm_{\ell,j} \Fm_{\ell,j}^\herm, 
	\end{align} 
	where $\zv_\ell^{\rm ul} \sim \Cc\Nc(0,1)$ is AWGN at RRH $\ell$.
	Taking the trace and dividing by $M$ we find the equivalent variance per component
	\begin{equation} 
		\sigma^2_\ell = \frac{1}{M} \trace ( \Xim_\ell ) =   1 + \SNR \left ( \sum_{j \neq \Uc_\ell}  \beta_{\ell,j} \right ).  \label{sigmaell}
	\end{equation}
	Under this assumption, we have that the LMMSE receiving vector is given by 
	\vspace{-.15cm}
	\begin{equation} 
		\vv_{\ell,k} = \left ( \sigma_\ell^2 \Id + \SNR \sum_{j \in \Uc_\ell} \hv_{\ell,j} \hv_{\ell,j}^\herm \right )^{-1} \hv_{\ell,k}.  \label{eq:lmmse}
	\end{equation}
	
	For the combining coefficients, we consider two options. The first one is equal gain combining (EGC), with $w_{\ell,k} = 1$ for all $\ell \in \Cc_k$. 
	The second option maximizes the SINR after combining.  
	The effective received signal model at RRH $\ell \in \Cc_k$ relative to UE $k$ can be written as
	\begin{equation} 
		r^{\rm ul}_{\ell,k} = \sqrt{\SNR} \left ( g_{\ell, k,k} s^{\rm ul}_k + \sum_{j \in \Uc_\ell: j \neq k} g_{\ell, k,j} s^{\rm ul}_j  \right ) +  \vv_{\ell, k}^\herm \xiv_\ell  \label{effective model} 
	\end{equation} 
	where we define
		$g_{\ell,k,j} = \vv_{\ell,k}^\herm \hv_{\ell,j}$
	and let $\xiv_\ell$ the unknown interference plus noise vector, assumed $\sim \Cc\Nc(\zerov, \sigma_\ell^2 \Id)$. 
	Stacking $\{r^{\rm ul}_{\ell,k} : \ell \in \Cc_k\}$ as a $|\Cc_k| \times 1$ column vector $\rv^{\rm ul}_k$, we can write the output symbols of cluster $\Cc_k$ relative to UE $k$ as
	\begin{equation}
		\rv^{\rm ul}_k = \sqrt{\SNR}  \left ( \av_k s^{\rm ul}_k  +  \Gm_k \sv_k^{\rm ul} \right ) + \zetav_k,
	\end{equation} 
	where 
		$\zetav_k = \{ \vv_{\ell, k}^\herm \xiv_\ell  : \ell \in \Cc_k \}$
	has the covariance matrix given by 
		$\Dm_k = \diag \left \{ \sigma_\ell^2 \|\vv_{\ell,k}\|^2 : \ell \in \Cc_k \right \}$
	and $\av_k = \{ g_{\ell, k,k} : \ell \in \Cc_k\}$.

	The matrix $\Gm_k \in \CC^{|\Cc_k| \times (|\Uc(\Cc_k)| - 1)}$ contains elements $g_{\ell, k, j}$ in position corresponding to 
	RRH $\ell$ and UE $j$ (after a suitable index reordering) if $(\ell,j) \in \Ec$, and zero elsewhere. The vector $\sv_k^{\rm ul} \in \CC^{(|\Uc(\Cc_k)| - 1) \times 1}$ contains the symbols of all users $j \in \Uc(\Cc_k) : j \neq k$. 
	Then, the total interference plus noise covariance matrix given the available CSI is
	\begin{equation}
		\Gammam_k = \Dm_k  + \SNR \; \Gm_k \Gm^\herm_k \label{eq:interference_plus_noise_cov}
	\end{equation}
	and the corresponding {\em nominal} SINR for user $k$ with combining is given by 
	\vspace{-0.2cm}
	\begin{equation} 
		\SINR^{\rm ul-nom}_k = \frac{\SNR \; \wv_k^\herm \av_k \av_k^\herm \wv_k}{\wv_k^\herm \Gammam_k  \wv_k}.  \label{SINRnom}
	\end{equation}
	The maximization of this nominal SINR with respect to $\wv_k$ amounts to 
	find the maximum generalized eigenvalue of the matrix pencil 
	$(\av_k \av_k^\herm, \Gammam_k )$. Since the matrix $\av_k \av_k^\herm$ has rank 1 and therefore it has only one non-zero eigenvalue, 
	the solution is readily given by 
	\vspace{-0.1cm}	
	\begin{equation}
		\wv_k = \Gammam_k^{-1} \av_k. \label{eq:w_opt}
	\end{equation}
For both MRC and LMMSE receive combining, the overall received vector is obtained by forming the vector $\overline{\vvv}_k$ by stacking the $|\Cc_k|$ blocks  of dimensions $M \times 1$ given by  $w_{\ell,k} \vv_{\ell,k}$ on top of each other and normalizing such that $\overline{\vvv}_k$ has unit norm. 
After expanding $\overline{\vvv}_k$ to $\vvv_k$ of dimension $LM \times 1$ by inserting the all-zero blocks corresponding to the RRHs $\ell \notin \Cc_k$, the resulting  SINR is again given by (\ref{UL-SINR-unitnorm}). This scheme differs from the distributed large-scale fading decoding (LSFD) in \cite{9336188}, as the LSFD relies on the expected channel vectors $\{\EE[\hv_{\ell,k}]  : (\ell, k) \in \Ec \}$. In contrast, we use the instantaneous channel realization estimate for the computation of the combining coefficients.
\vspace{-.1cm}
\vspace{-.05cm}
\section{UL channel estimation}\vspace{-.05cm}
In practice, ideal (although partial) CSI is not available and the channels $\{\hv_{\ell,k} : (\ell,k) \in \Ec\}$ must be estimated from
the UL pilots.  The pilot field received at RRH $\ell$ is given by the $M \times \tau_p$ matrix of received symbols
$\Ym_\ell^{\rm pilot} = \sum_{i=1}^K \hv_{\ell,i} \phiv_{t_i}^\herm + \Zm_\ell^{\rm pilot}$, 
where $\phiv_{t_i}$ denotes the pilot vector of dimension $\tau_p$ used by UE $i$ in the current RB, with  
total energy $\| \phiv_{t_i} \|^2 = \tau_p \SNR$.  For each UE $k \in \Uc_\ell$, RRH $\ell$ produces the {\em pilot matching} (PM) channel estimates
\begin{eqnarray} 
	\widehat{\hv}^{\rm pm}_{\ell,k} & = & \frac{1}{\tau_p \SNR} \Ym^{\rm pilot}_\ell \phiv_{t_k}  \\
	& = & \hv_{\ell,k}  + \sum_{\substack{i : t_i = t_k \\ i\neq k}} \hv_{\ell,i}  + \widetilde{\zv}_{t_k,\ell} ,  \label{chest}
\end{eqnarray} 
where $\widetilde{\zv}_{t_k,\ell}$ has i.i.d. with components $\Cc\Nc(0, \frac{1}{\tau_p\SNR})$. 
Notice that the presence of UEs $i \neq k$ using the same pilot $t_k$ yields pilot contamination. 

Assuming that the subspace information $\Fm_{\ell,k}$ of all $k \in \Uc_\ell$ is known, 
we consider also the {\em subspace projection} (SP)  pilot decontamination scheme for which the projected channel estimate is given by the 
orthogonal projection of $\widehat{\hv}^{\rm pm}_{\ell,k}$ onto the subspace spanned by the columns of $\Fm_{\ell,k}$, i.e., 
\begin{align}
	\widehat{\hv}^{\rm sp}_{\ell,k} &= \Fm_{\ell,k}\Fm_{\ell,k}^\herm \widehat{\hv}^{\rm pm}_{\ell,k} 
	\label{chest1}
\end{align}
The pilot contamination term after the subspace projection  
is a Gaussian vector with mean zero and covariance matrix
\begin{equation}
	\Sigmam_{\ell,k}^{\rm co}  = \sum_{\substack{i : t_i = t_k \\ i\neq k}} \frac{\beta_{\ell,i} M}{|\Sc_{\ell,i}|}  \Fm_{\ell,k} \Fm_{\ell,k}^\herm \Fm_{\ell,i} \Fm^\herm_{\ell,i} \Fm_{\ell,k} \Fm^\herm_{\ell,k}. 
\end{equation}
From this expression it can be argued that when $\Fm_{\ell,k}$ and $\Fm_{\ell,i}$ are nearly mutually orthogonal, i.e. $\Fm_{\ell,k}^\herm \Fm_{\ell.i} \approx \zerov$,
the subspace projection is able to significantly reduce the pilot contamination effect.

The procedure for the signal detection relying on UL pilot-based channel estimates is similar to the one with ideal partial CSI, by simply replacing 
the ideal partial CSI $\{ \hv_{\ell,k} : (\ell,k) \in \Ec\}$ with the estimated partial CSI $\{\widehat{\hv}_{\ell,k} : (\ell,k) \in \Ec\}$, 
where $\widehat{\hv}_{\ell,k} = \widehat{\hv}^{\rm pm}_{\ell,k}$ or $\widehat{\hv}_{\ell,k} = \widehat{\hv}^{\rm sp}_{\ell,k}$ dependent on the channel estimation method.  
Here we only assume that the variance of the unknown interference plus noise term $\sigma_\ell^2$ defined in (\ref{sigmaell}) is  known, 
since this depends on  aggregate average signal power, which is only a function of the LSFCs and therefore of the system geometry. 
Since after these substitutions the expressions of the receiver vectors are identical as before, they shall not be repeated here.


\section{Simulations and concluding remarks} 


We evaluate the performance of the different receiver combining schemes for ideal partial and estimated CSI. 
In our simulations, we consider a coverage area of $A = 500\times 500$ meters with a torus topology to avoid boundary effects. 
The LSFCs are given  according to the 3GPP urban microcell pathloss model from \cite{3gpp38901}. 
The parameter $\SNR$ is chosen such that $\beta_{\ell,k} M \SNR = 1$ (i.e., 0 dB), when $\beta_{\ell,k}$ is calculated for distance 
$3 d_L$, where $d_L = 2 \sqrt{\frac{A}{\pi L}}$ is the diameter of a disk of area equal to $A/L$. 
We consider RBs of dimension $T = 200$ symbols, consistently with the 14 OFDM symbols $\times$ 12 subcarriers specified in LTE and 5GNR. 
The spectral efficiency for UE $k$ is given by  ${\rm SE}^{\rm ul}_k =  (1 - \tau_p/T) R_k^{\rm ul}$, where
the UL pilot redundancy factor multiplies the optimistic ergodic rate defined in (\ref{ergodic-rate}).
The angular support $\Sc_{\ell,k}$ contains the DFT quantized angles (multiples of $2\pi/M$) falling inside an interval of length $\Delta$ placed symmetrically around the direction joining UE $k$ and RRH $\ell$. If not otherwise mentioned, in the simulated system $K = 100$ UEs are served by $L = 50$ RRHs, each with $M=64$ antennas, and we use $\Delta = \pi/16$, the QoS threshold $\eta=1$ and the maximum cluster size $Q=30$.
For each set of parameters we generated 48 independent layouts (random uniform placement of RRHs and UEs and greedy cluster formation), 
and for each  layout we computed the expectation in (\ref{ergodic-rate}) by Monte Carlo averaging with respect to the channel vectors. 
In all figures, ``full CSI'' indicates ideal partial CSI as discussed before. 

\begin{figure}[t]
	\centerline{\includegraphics[width=.45\linewidth]{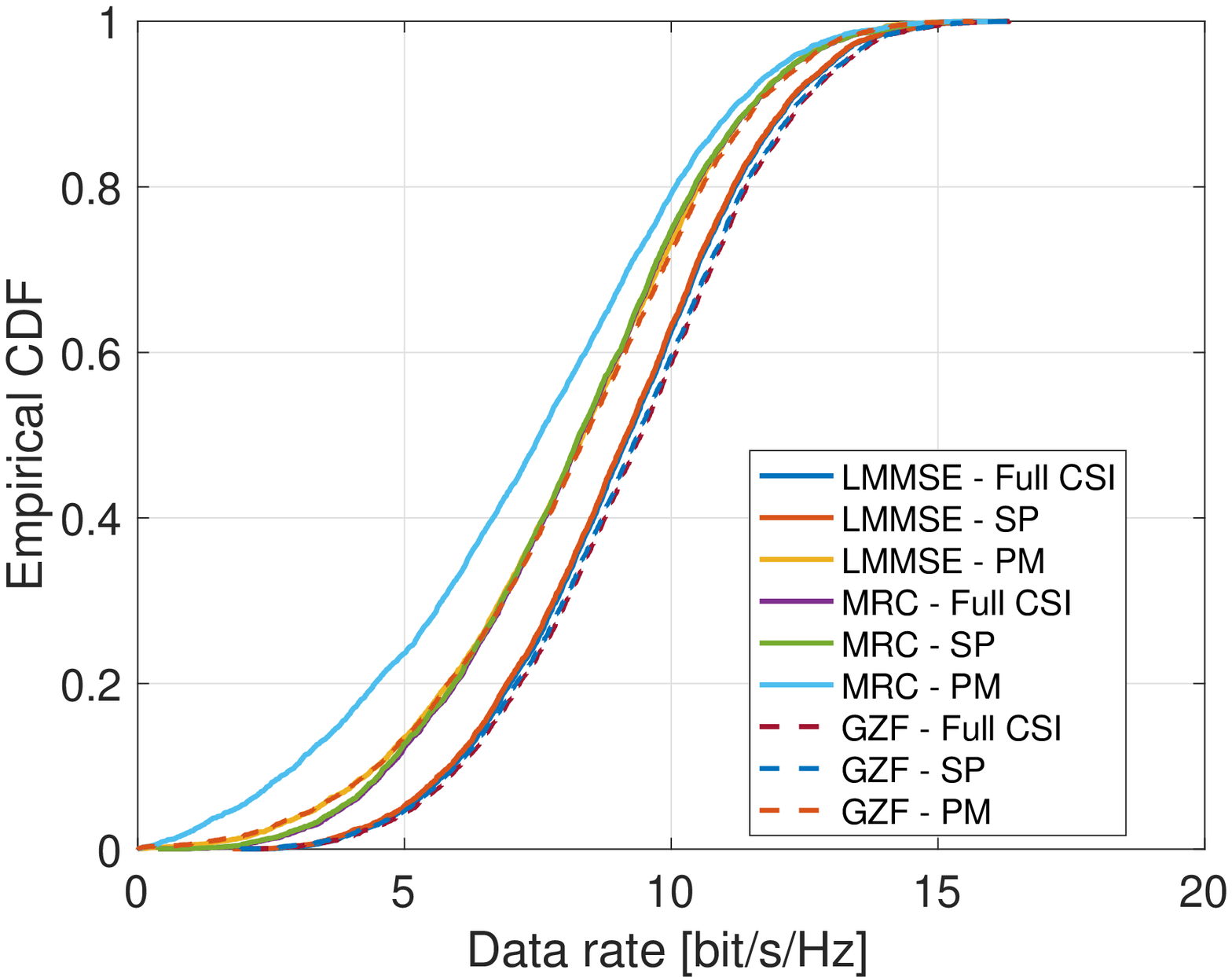} \hspace{.05cm} \includegraphics[trim=20 0 40 25, clip,width=.45\linewidth]{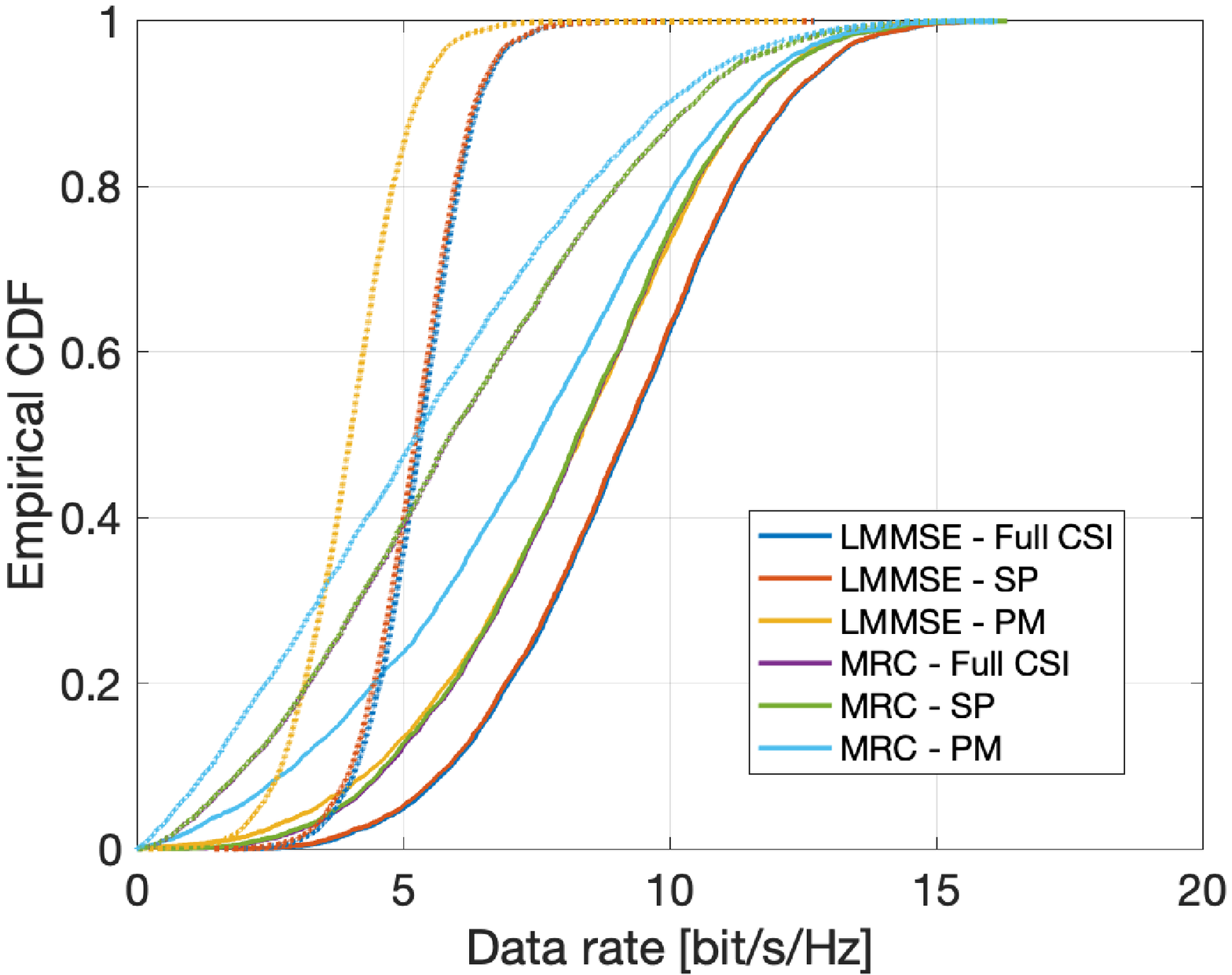}}
	\vspace{-0.3cm}\caption{Empirical CDF of the UL data rate per UE with $\tau_p=20$. Left: results for GZF and for local MRC and MMSE with global optimized combining. Right:  results for local MRC and MMSE with global optimized (solid lines) and equal gain (dotted lines) combining.}
	\label{fig:se_combining_taup20}
	\vspace{.1cm}
	\centerline{\includegraphics[width=.45\linewidth]{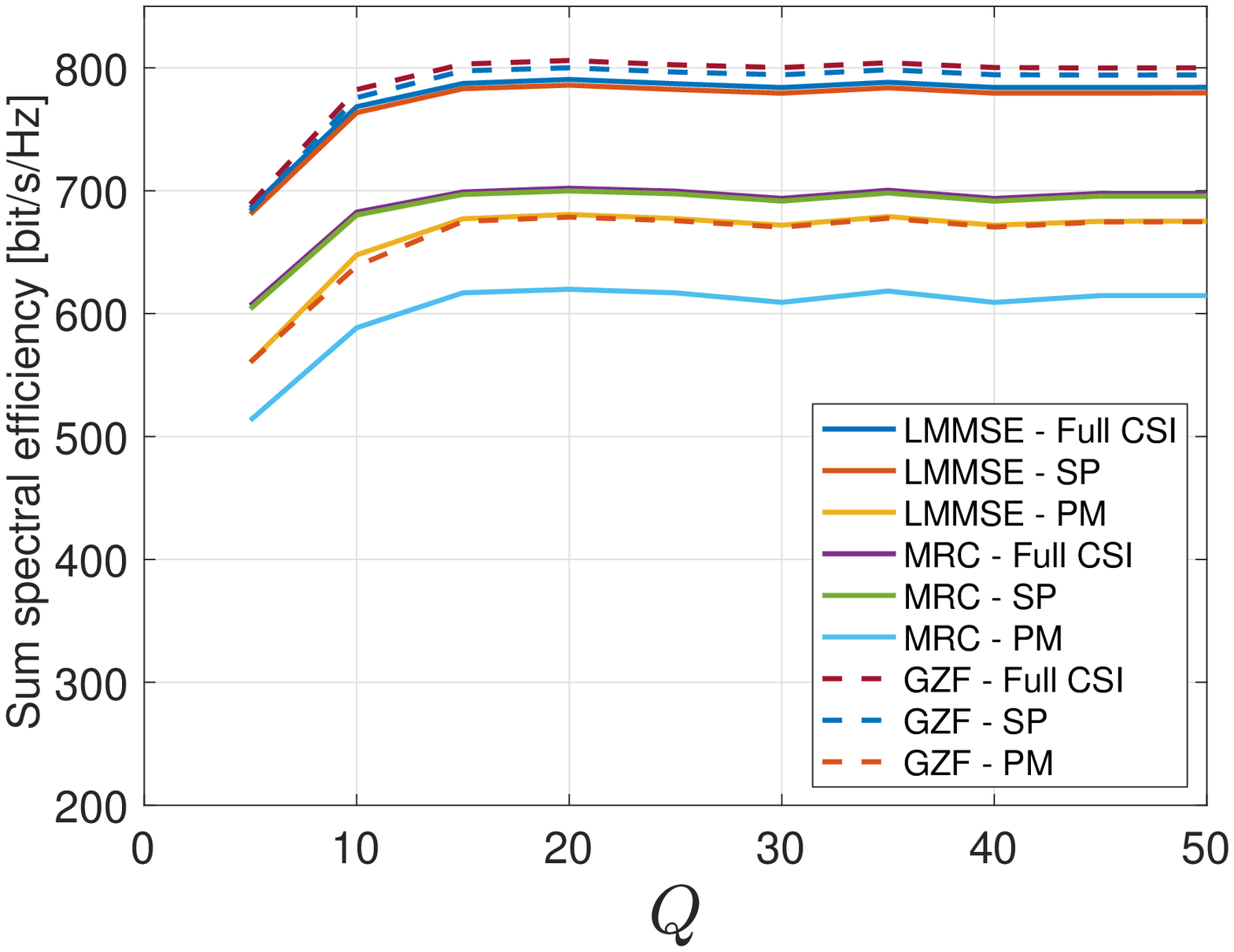} \hspace{.05cm} \includegraphics[width=.45\linewidth]{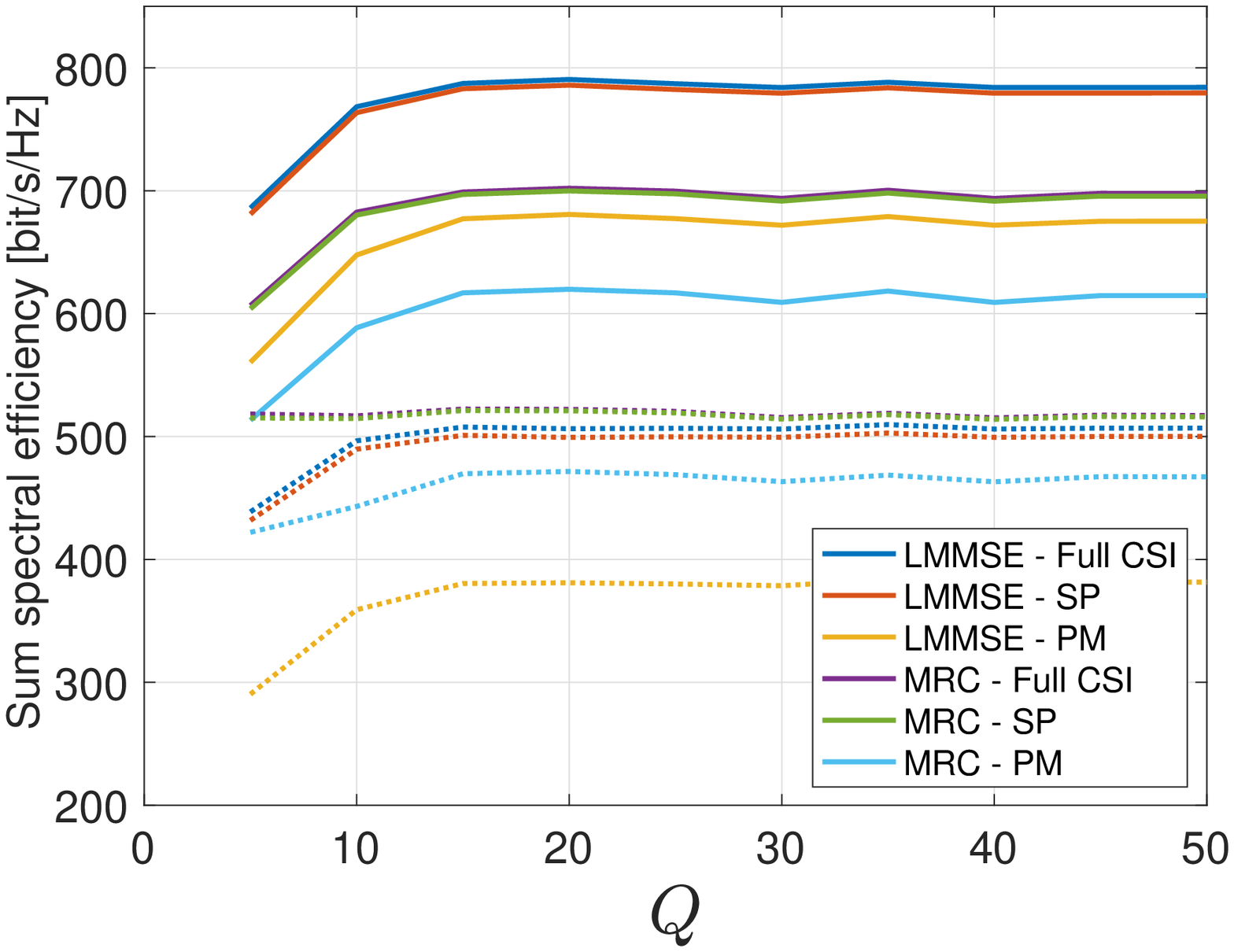}}
	\vspace{-0.3cm}\caption{UL sum spectral efficiency for different maximum cluster sizes $Q$. Left: results for GZF and for local MRC and MMSE with global optimized combining. Right: results for local MRC and MMSE with optimized (solid lines) and equal gain (dotted lines) combining.}
	\label{fig:se_vs_Q}
\end{figure}

Fig.~\ref{fig:se_combining_taup20} shows the CDF of the optimistic ergodic rate per UE for $\tau_p = 20$ for the different receiver schemes. 
GZF outperforms the local MRC and LMMSE combining, as expected. For local scheme with global combining, 
LMMSE outperforms MRC, as it takes care of the most significant interference components.  In general, optimized combining 
achieve a significant performance improvement with respect to EGC. 
Notice also that the difference between ideal partial CSI and SP pilot-based  estimation is quite small. 
In contrast, PM pilot-based estimation incurs a significant performance loss. 
This indicates that SP is able to remove a significant part of the pilot contamination and that the latter has a significant impact if not properly handled. 
In Fig.~\ref{fig:se_vs_Q}, we see that $Q = 15$ already saturates the sum SE performance. Therefore, enlarging further the cluster size yields only a complexity increase without return in performance. In contrast, as $Q$ drops below 10, the sum SE degrades significantly. 

\begin{figure}[h]
	\centerline{\includegraphics[width=.45\linewidth]{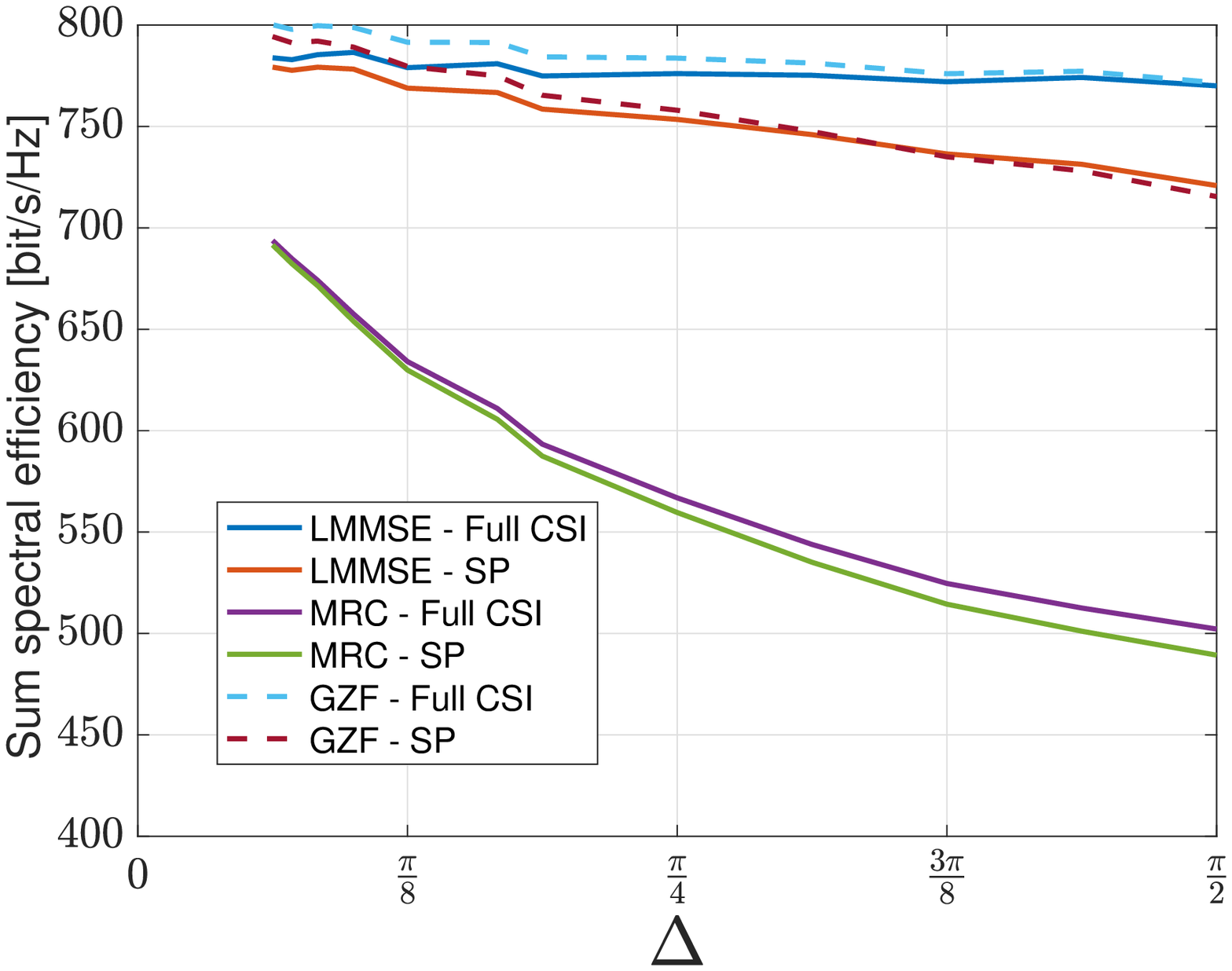} \hspace{.05cm} \includegraphics[width=.45\linewidth]{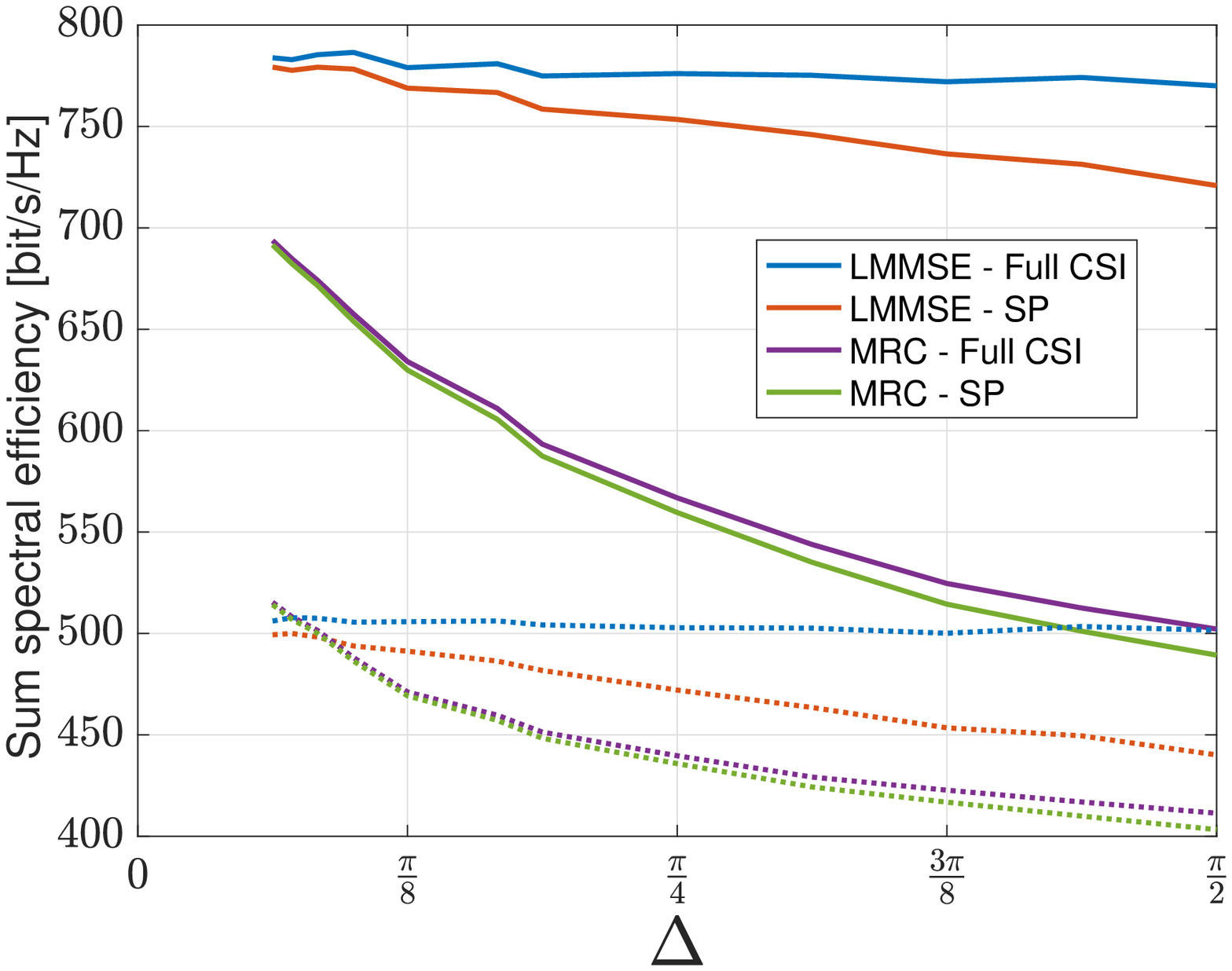}}
	\vspace{-0.3cm}\caption{UL sum spectral efficiency compared to the angular spread $\Delta$. Left: results for GZF and for local MRC and MMSE with global optimized combining. Right: results for local MRC and MMSE with optimized (solid lines) and equal gain (dotted lines) combining.}
	\label{fig:se_vs_delta}
\end{figure}

Fig.~\ref{fig:se_vs_delta} shows that, as expected, the performance of the SP degrades for larger values of $\Delta$, as the channels become more 
isotropic and SP is less effective in reducing pilot contamination. However, the degradation w.r.t. ideal CSI is not dramatic. 
For $\Delta=\frac{\pi}{16}$, it is less than $1.5 \%$ for all receiver schemes, whereas it becomes approximately $2 \%$ (MRC) and 
between $6 \%$ and $12 \%$ (GZF and LMMSE combining) for $\frac{\pi}{2}$.

\begin{figure}[t]
	\centerline{\includegraphics[width=.45\linewidth]{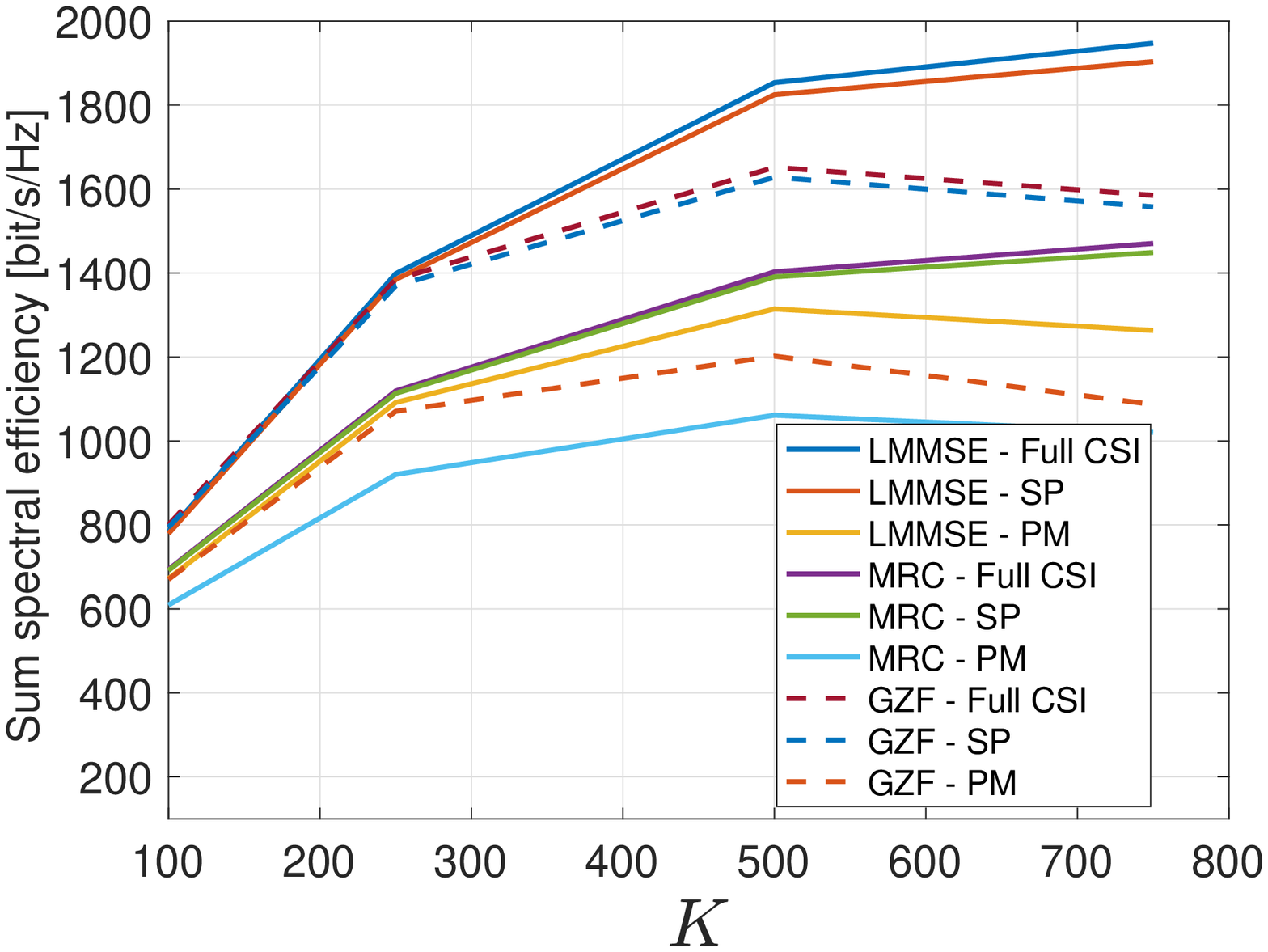} \hspace{.05cm} \includegraphics[width=.45\linewidth]{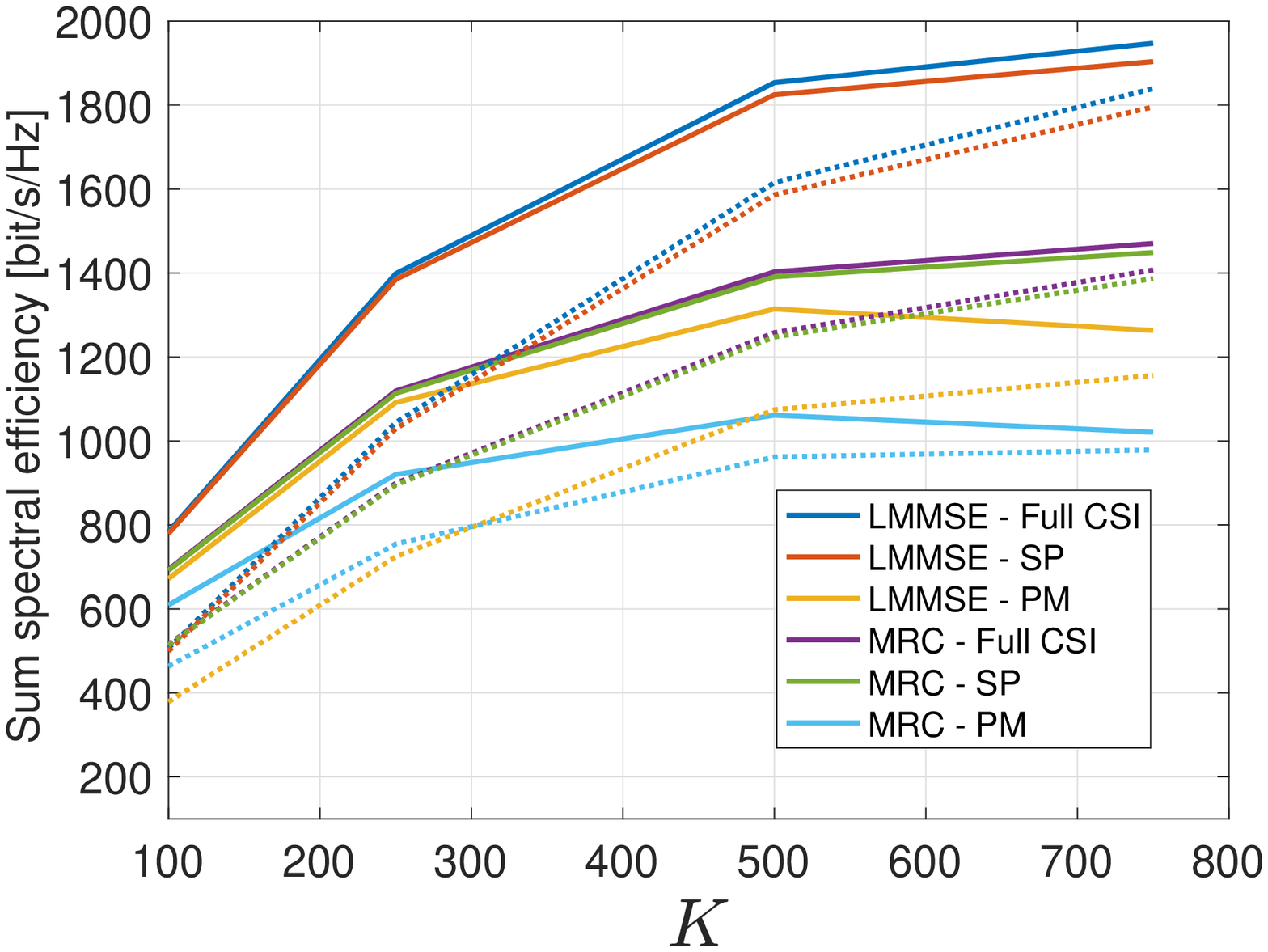}}
	\vspace{-0.3cm}\caption{UL sum spectral efficiency compared to the number of UEs $K$. Left: results for GZF and for local MRC and MMSE with global optimized combining. Right: results for local MRC and MMSE with optimized (solid lines) and equal gain (dotted lines) combining.}
	\label{fig:se_vs_K}
	\vspace{.1cm}
	\centerline{\includegraphics[width=.45\linewidth]{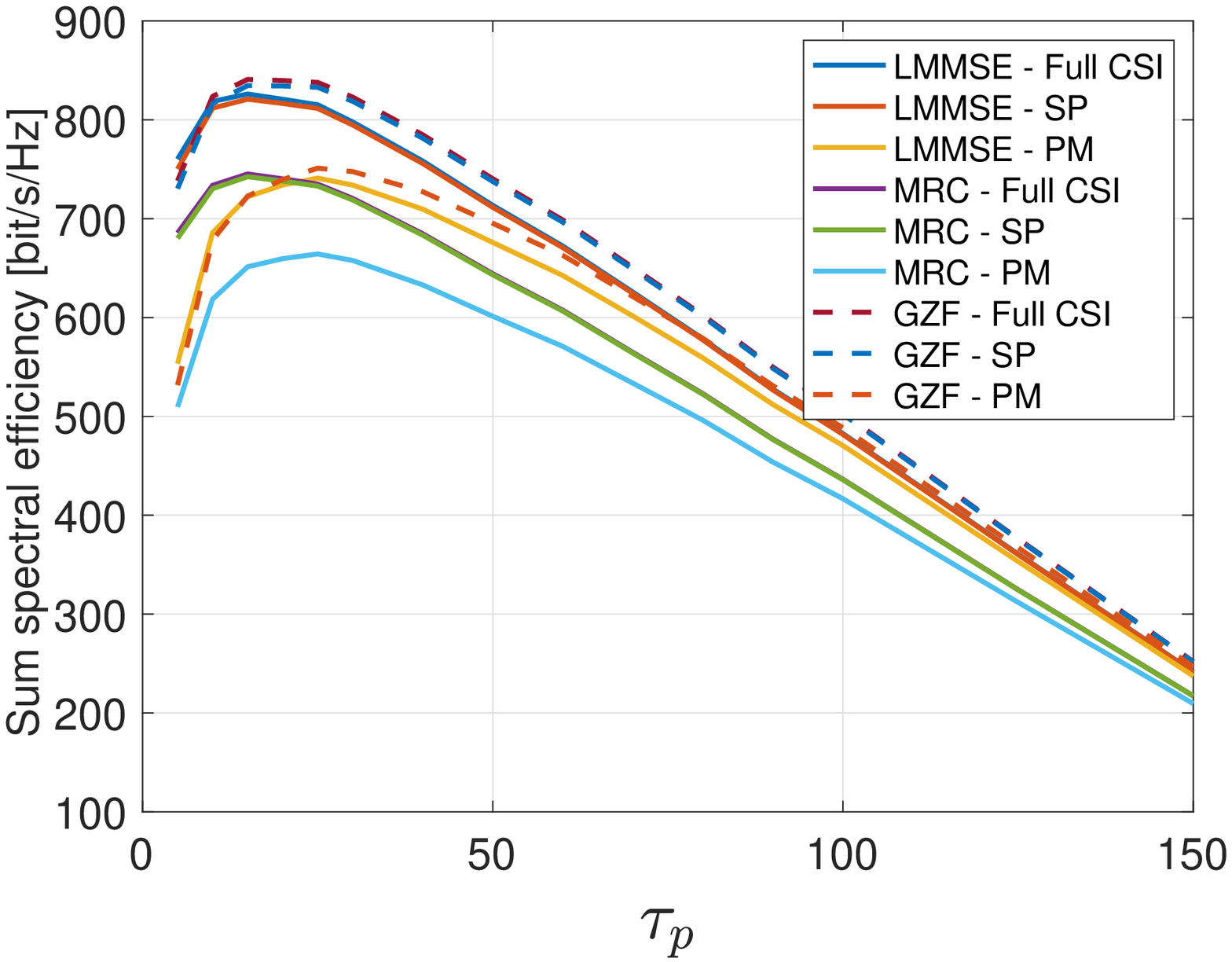} \hspace{.05cm} \includegraphics[width=.45\linewidth]{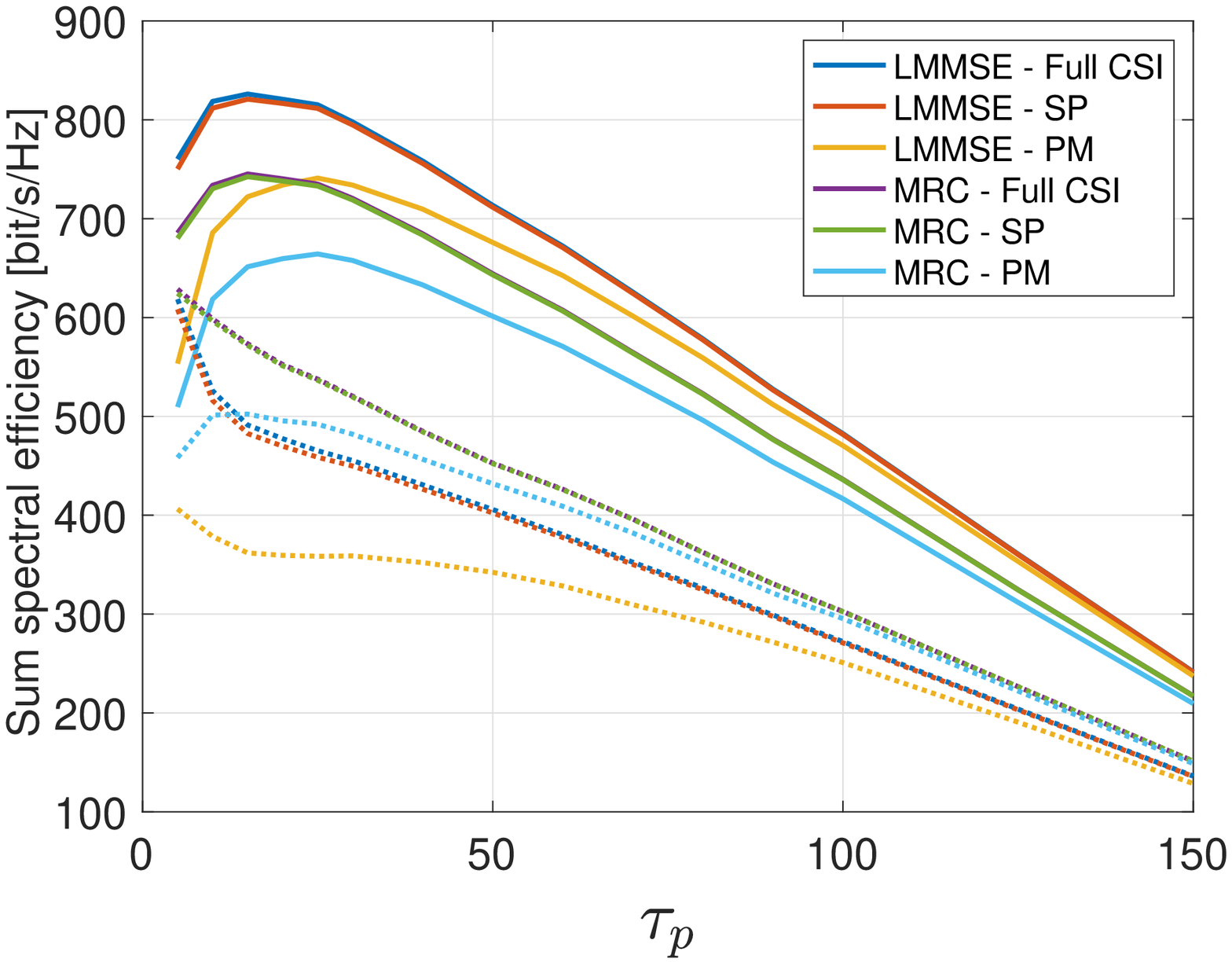}}
	\vspace{-0.3cm}\caption{UL sum spectral efficiency for different values of $\tau_p$. Left: results for GZF and for local MRC and MMSE with global optimized combining. 
	Right: results for local MRC and MMSE with optimized (solid lines) and equal gain (dotted lines) combining.}
	\label{fig:se_vs_tau_p}
\end{figure}

If we increase the number of UEs, we can observe in Fig.~\ref{fig:se_vs_K} that the sum SE increases as well until $K=500$. For $K=750$, the results differ by the receiver scheme. For LMMSE/MRC with ideal partial CSI and SP, the sum SE keeps increasing. For GZF, as well as for LMMSE/MRC with optimized gains and PM, the sum SE decreases. This is because of the increasing effect of pilot contamination for PM and because the information becomes too partial for GZF. For all receive combining schemes, the gap between optimized gains and EGC decreases with more UEs. In all scenarios, each UE could be served by 
at least one RRH, i.e., there is no outage.

Fig.~\ref{fig:se_vs_tau_p} shows the sum SE for different values of $\tau_p$. The tradeoff between quality of the CSI estimation and pilot redundancy is evident for
the best schemes (GZF, LMMSE and MRC with optimized combining), where the SE curves exhibit a maximum for $\tau_p$ between 20 and 30. 
In contrast, LMMSE and MRC with EGC have an almost monotonically decreasing behavior with respect to $\tau_p$. We interpret this fact as follows: these schemes perform quite poorly in terms of multiuser MIMO beamforming and interference suppression in the spatial domain, and therefore do not take advantage 
of a high quality CSI in exchange of a larger pilot redundancy.

\bibliography{IEEEabrv,spawc-paper}

\begin{thebibliography}{1}
\providecommand{\url}[1]{#1}
\csname url@samestyle\endcsname
\providecommand{\newblock}{\relax}
\providecommand{\bibinfo}[2]{#2}
\providecommand{\BIBentrySTDinterwordspacing}{\spaceskip=0pt\relax}
\providecommand{\BIBentryALTinterwordstretchfactor}{4}
\providecommand{\BIBentryALTinterwordspacing}{\spaceskip=\fontdimen2\font plus
\BIBentryALTinterwordstretchfactor\fontdimen3\font minus
  \fontdimen4\font\relax}
\providecommand{\BIBforeignlanguage}[2]{{%
\expandafter\ifx\csname l@#1\endcsname\relax
\typeout{** WARNING: IEEEtran.bst: No hyphenation pattern has been}%
\typeout{** loaded for the language `#1'. Using the pattern for}%
\typeout{** the default language instead.}%
\else
\language=\csname l@#1\endcsname
\fi
#2}}
\providecommand{\BIBdecl}{\relax}
\BIBdecl

\bibitem{caire2003achievable}
G.~Caire and S.~Shamai, ``On the achievable throughput of a multiantenna
  gaussian broadcast channel,'' \emph{{IEEE Trans. on Inform. Theory}},
  vol.~49, no.~7, pp. 1691--1706, 2003.

\bibitem{3gpp38211}
3GPP, ``{Physical channels and modulation (Release 16)},'' 3GPP Technical
  Specification 38.211, 12 2020, {Version 16.4.0}.

\bibitem{marzetta2010noncooperative}
T.~L. Marzetta, ``Noncooperative cellular wireless with unlimited numbers of
  base station antennas,'' \emph{{IEEE Trans. on Wireless Comm.}}, vol.~9,
  no.~11, pp. 3590--3600, 2010.

\bibitem{wyner1994shannon}
A.~D. Wyner, ``Shannon-theoretic approach to a gaussian cellular
  multiple-access channel,'' \emph{{IEEE Trans. on Inform. Theory}}, vol.~40,
  no.~6, pp. 1713--1727, 1994.

\bibitem{9336188}
{\"O}.~T. Demir, E.~Bj{\"o}rnson, L.~Sanguinetti \emph{et~al.}, ``{Foundations
  of User-Centric Cell-Free Massive MIMO},'' \emph{Foundations and
  Trends{\textregistered} in Signal Processing}, vol.~14, no. 3-4, pp.
  162--472, 2021.

\bibitem{9064545}
E.~Bj{\"o}rnson and L.~Sanguinetti, ``{Scalable Cell-Free Massive MIMO
  Systems},'' \emph{{IEEE Trans. on Comm.}}, vol.~68, no.~7, pp. 4247--4261,
  2020.

\bibitem{adhikary2013joint}
A.~Adhikary, J.~Nam, J.-Y. Ahn, and G.~Caire, ``{Joint Spatial Division and
  Multiplexing—The Large-Scale Array Regime},'' \emph{{IEEE Trans. on Inform.
  Theory}}, vol.~59, no.~10, pp. 6441--6463, 2013.

\bibitem{marzetta2016fundamentals}
T.~L. Marzetta, E.~G. Larsson, H.~Yang, and H.~Q. Ngo, \emph{{Fundamentals of
  Massive MIMO}}.\hskip 1em plus 0.5em minus 0.4em\relax Cambridge University
  Press, 2016.

\bibitem{3gpp38901}
3GPP, ``{Study on channel model for frequencies from 0.5 to 100 GHz (Release
  16)},'' 3GPP Technical Specification 38.901, 12 2019, {Version 16.1.0}.

\end{thebibliography}

\end{document}